\pdfoutput=1
\documentclass[runningheads]{llncs}

\usepackage[utf8]{inputenc} 
\usepackage[T1]{fontenc}    
\usepackage{color}
\usepackage{hyperref}       
\usepackage{url}            
\usepackage{makecell}
\usepackage{booktabs}       
\usepackage{amsfonts}       
\usepackage{nicefrac}       
\usepackage{microtype}      
\usepackage{lipsum}
\usepackage{graphicx}
\usepackage{ragged2e}
\usepackage[font=small,labelfont=bf]{caption}
\graphicspath{ {./images/} }
\def\etal{\textit{et al. }}

\title{Topology-Preserving Segmentation Network: A Deep Learning Segmentation Framework for Connected Component}

\author{
 Han Zhang\inst{1} \and Lok Ming Lui\inst{1}
}
\institute{Dept. of Mathematics, The Chinese University of Hong Kong,\\ Hong Kong, China}
\begin{document}
\maketitle

\begin{abstract}
Medical image segmentation, which aims to automatically extract anatomical or pathological structures, plays a key role in computer-aided diagnosis and disease analysis. Despite the problem has been widely studied, existing methods are prone to topological errors. In medical imaging, the topology of the structure, such as the kidney or lung, is usually known. Preserving the topology of the structure in the segmentation process is of utmost importance for accurate image analysis. In this work, a novel learning-based segmentation model is proposed. A {\it topology-preserving segmentation network (TPSN)} is trained to give an accurate segmentation result of an input image that preserves the prescribed topology. TPSN is a deformation-based model that yields a deformation map through a UNet, which takes the medical image and a template mask as inputs. The main idea is to deform a template mask describing the prescribed topology by a diffeomorphism to segment the object in the image. The topology of the shape in the template mask is well preserved under the diffeomorphic map. The diffeomorphic property of the map is controlled by introducing a regularization term related to the Jacobian in the loss function. As such, a topology-preserving segmentation result can be guaranteed. Furthermore, a multi-scale TPSN is developed in this paper that incorporates multi-level information of images to produce more precise segmentation results. To evaluate our method, we applied the 2D TPSN on Ham10000 and 3D TPSN on KiTS21. Experimental results illustrate our method outperforms the baseline UNet segmentation model with/without connected-component analysis (CCA) by both the dice score and IoU score. Besides, results show that our method can produce reliable results even in challenging cases, where pixel-wise segmentation models by UNet and CCA fail to obtain accurate results.


\end{abstract}


\section{Introduction}

Automated anatomical segmentation is an important procedure for many clinical applications, such as diagnosing diseases and monitoring disease progression. Recent developments in deep learning have achieved great success in medical image segmentation. Once successfully trained, the deep neural network can segment the input image in real-time. Nevertheless, the major challenge is the accuracy of the segmentation result. Accurate segmentation is essential for meaningful medical image analysis. In particular, the accurate topology of the segmented anatomical structure is crucial for the measurement of the structure. Unfortunately, most existing algorithms are prone to topological errors, causing mis-segmentation with multiple components or thin connections. In practical situations, the topologies of anatomical structures are often known. For instance, structures like the kidney or liver are known to be simply-connected. We are thus motivated to develop a deep segmentation model with a topological prior that learns to segment with correct topology. State-of-the-art deep segmentation approaches commonly assign a label to each pixel and can be formulated as a pixel-wise classification problem. Under this formulation, the topological constraint is hard to be enforced. Deformation-based models\cite{chan2018topology,siu2020image,zhang2021topoconv,zhang2021topology}, which deform a template mask by a suitable deformation map to segment the image, have been recently proposed. The topological prior can be easily incorporated using this approach, which will be adopted in this work.

In this paper, a learning-based topology-preserving segmentation framework that learns to segment an input image with a given topological prior is proposed. TPSN takes the image to be segmented and a template mask describing the prescribed topology as inputs. It outputs a diffeomorphic deformation map that deforms the template mask to segment the input image. For instance, suppose the kidney in an image is to be extracted, the template shape can be chosen as a simple disk, and will then be deformed by the diffeomorphism to enclose the kidney in the input image. The diffeomorphic deformation map ensures the topology of the segmented shape is consistent with that of the template. The diffeomorphic property of the map is enforced by regularizing the Jacobian and the Laplacian of the map. 
With the trained network, the segmentation process can be done efficiently and the topology of the segmented shape is guaranteed to be consistent with the prescribed topological prior. To further improve the segmentation result, a {\it multi-level topology-preserving segmentation network (mlTPSN)}, which incorporates multi-level information by producing masks from coarse to fine, is introduced. At the low-fidelity level, a low-resolution version of the image is fitted into the network to predict a rough mask. In the higher levels, rough masks are used as the template mask to generate a more precise mask to segment the higher-resolution image. This multi-level strategy is proved to be effective to enhance the accuracy of the segmentation result. To evaluate the proposed TPSN and mlTPSN, experiments are carried out on both 2D and 3D medical images. Our methods outperform the baseline UNet with and without connected component analysis. In challenging situations when the UNet segmentation fails, our proposed networks can still yield satisfying and reliable results. These demonstrate the efficacy of our proposed models.

Our contribution of this paper is three-fold:
(1) We introduced a deep-learning segmentation network that learns to segment an image accurately with correct topology according to the prescribed topological prior.
(2) We designed a special regularization, namely the {\it $\epsilon$-ReLU Jacobian loss}, which can effectively enforce the bijectivity of the deformation map and prevent heavy shrinkage.
(3) We proposed a multi-scale topology-preserving segmentation network that incorporates multi-level information of images to produce more precise results.

\section{Related Work}
Medical image segmentation has been widely studied. Numerous deep learning based models have been recently proposed. UNet and its variants achieve great success\cite{ronneberger2015u,milletari2016v,cciccek20163d}. In particular, 3D-UNet \cite{cciccek20163d} and VNet \cite{milletari2016v} proposed by Çiçek\etal and Milletari\etal respectively are successful for 3D medical image segmentation. In the challenge of KiTS19\cite{heller2021state}, nnUNet \cite{isensee2021nnu} has achieved outstanding results using UNet and a series of practical pre- and post- processing strategies. Besides, \cite{jha2020doubleu} proposed to use one UNet to predict an initial mask and apply an addtional one to refine the segmentation result.

Mathematical models for image segmentation have also been extesnively studied. \cite{kass1988snakes} proposed an active contour model that delineates the object boundary. Chan and Vese\cite{chan2001active} generalized active contour model using a level set formulation. Shape prior segmentation have been investigated. Segmentation with topological prior has shown to be effective in enhanching the accurary \cite{heller2021state}. Chan \etal \cite{chan2018topology} introduced a deformation based segmenation model using quasiconformal maps for topology preserving segmentation. Siu \etal \cite{siu2020image} introduced to incorporate the dihedral angle in the model for segmentation with partial convexity prior and topological prior. Zhang \etal \cite{zhang2021topology} proposed a deformation based topology-preserving segmentation model through registration using the hyperelastic regularization. The model has been further extended to incorporate the convexity prior\cite{zhang2021topoconv}. Zhou \etal \cite{zhou2019prior} propsoed a model, which accounts for the relative location and size prior statistically. Hu \etal \cite{hu2019topology} designed a continuous-valued loss function to enforce the topological constraint. Shit \etal \cite{shit2021cldice} introduced clDice to enhance the connectedness in the segmentation for tubular structures. Wyburd \etal \cite{wyburd2021teds} proposed a topology-preserving deep segmentation network by compositing a series of bijection deformations.

\section{Method}
\begin{figure}
    \centering
    \includegraphics[width = 0.95\textwidth]{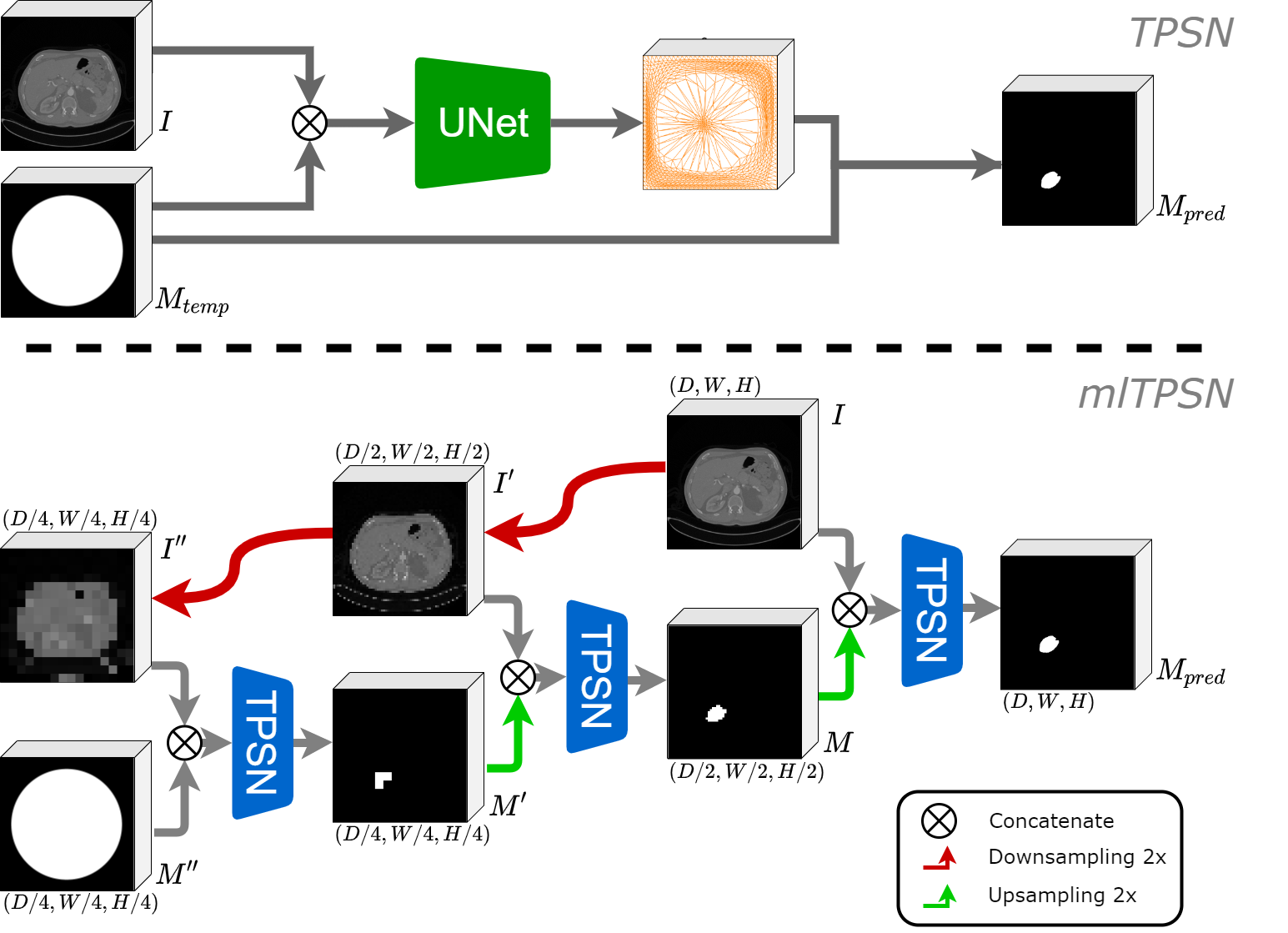}
    \caption{The framework and architecture for TPSN and multi-level TPSN}
    \label{fig:framework}
\end{figure}

In this section, we describe our proposed learning-based segmentation framework in detail. Conventional segmentation approaches are commonly formulated as a pixel-wise classification problem, with which the topological constraint is hard to be integrated. In this work, our strategy is to train a deep neural network to transform a template mask capturing the prescribed topological information by a diffeomorphism to segment the input image.



\smallskip

\noindent\textbf{Network Architecture}
The pipeline and architecture of our proposed model are illustrated in Figure \ref{fig:framework}. The top diagram shows our proposed {\it Topology-preserving Segmentation Network (TPSN)}. TPSN takes the image $I$ to be segmented and the template masks $M_{temp}$ as the inputs. The template mask is a binary image describing the prior topological information. For instance, if the structure to be extracted is simply-connected, $M_{temp}$ can be chosen as a binary image containing a 2D disk or 3D ball for 2D and 3D segmentation respectively. The network architecture is a UNet. It outputs the deformation mapping $f$. This output can be regarded as a 2-channel or 3-channel image capturing the coordinates of the deformation mapping for 2D or 3D segmentation respectively. $M_{temp}$ is then deformed by $f$ to obtain a transformed mask $M_{pred}$, which segments the structure in the image $I$.


\smallskip

\noindent\textbf{The Loss function}
To train the network, a suitable loss function is necessary. The loss function $\mathcal{L}$ aims to enforce the diffeomorphic property of the output deformation map, as well as ensuring that $M_{pred}$ is close to the ground truth segmentation result from the training data. $\mathcal{L}$ consists of the fidelity term and regularization term. The fidelity term is chosen as the Dice loss that drives $M_{pred}$ to the ground truth. To enforce the diffeomorphic property, regularization of the Jacobian is performed. A map is diffeomorphic if its Jacobian is positive everywhere \cite{zhang2021topology}. The following $\epsilon$-RELU Jacobian regularizer is applied:

\begin{equation}
\mathcal{L}_{Jac}(f) = || ReLU_{\epsilon} ( - \det\nabla(f)) ||_1
\end{equation}
where $ReLU_{\epsilon}(x) = Relu(x + \epsilon)$. $\mathcal{L}_{Jac}$ promotes a positive Jacobian determinant of $f$ that is greater than $\epsilon$ everywhere. This prevents flips (non-bijectivity) and heavy shrinkage of $f$. When $\epsilon$ is small, larger deformation can be acheived but more squeezing can be observed. When fine details are to be extracted, $\epsilon$ can be tuned to a smaller value. To further enhance the smoothness of $f$, the Laplacian loss is adopted:
\begin{equation}
\mathcal{L}_{Lap}(f) = || \Delta(f) ||_1
\end{equation}
Then, the final overall loss is as:
\begin{equation}
\mathcal{L} = \lambda_{Dice} \mathcal{L}_{Jac}(f)_{Dice}(M_{output},M_{label})
    +\lambda_{Jac.} \mathcal{L}_{Jac}(f)
    +\lambda_{Lap.} \mathcal{L}_{Lap}(f)
    \label{eq:loss}
\end{equation}
where $\mathcal{L}_{Dice}$,$\det\nabla(f)$ and $\Delta(f)$ are Dice loss, the Jacobian determinant of $f$ and Laplacian of $f$ weighted by $\lambda_{Dice}$, $\lambda_{Jac.}$ and $\lambda_{Lap.}$ respectively. The derivatives in the Jacobian and Laplacian are computed by forwarding difference and central difference schemes respectively.

\smallskip

\noindent\textbf{Multi-level TPSN}
To enhance the robustness and accuracy of our proposed model, a multi-level strategy is adopted to build the {\it multi-level topology-preserving segmentation network (ml-TPSN)}. The pipeline and architecture of ml-TPSN are shown in Figure \ref{fig:framework} bottom. The original image $I$ is downsampled into $I^{'}$ and $I^{''}$, which are respectively 1/2 and 1/4 of the original dimension. $I^{''}$ is fitted into TPSN with a coarse template mask of the same dimension as $I^{''}$ to output a predicted segmentation mask. The predicted segmentation mask gives a rough approximation of the segmentation. This coarse predicted mask is upsampled and used as the template mask $M^{'}$ for $I^{'}$. $I^{'}$ and $M^{'}$ are fitted into another TPSN to output the predicted segmentation mask for $I^{'}$.  The predicted mask is upsampled to $M$, which is fitted into a TPSN with $I$ to obtain the final segmentation result.

Such a multi-level strategy allows our model to gradually extract features from the image from coarse to fine. A predicted mask is used as the template mask in the next TPSN layer, which is closer to the actual mask. As a result, the algorithm is more robust, especially for structures with complex geometry that are to be segmented. Experiment results show that this multi-level strategy helps to improve the segmentation accuracy with higher segmentation scores.



\section{Experiment}

\subsection{Implementation Details}
{ The implementation details are described in this section. We compare our methods with the baseline UNet, as well as the UNet model with connected component analysis (CCA)\cite{heller2021state}\cite{isensee2021nnu}} For UNet with CCA, the simply-connected component with the largest area is chosen as the segmentation result to obtain a simply-connected segmentation result. The UNet model with connected component analysis is denoted by UNet(cca). The experiments of TEDSNet are carried out using the released code of \cite{wyburd2021teds}. 

\smallskip

\noindent\textbf{Dataset}
We evaluate TPSN and mlTPSN on two datasets, namely, Ham10000\cite{tschandl2018ham10000,tschandl2020human} and KiTS21\cite{heller2021state}. Most of the masks provided in Ham10000 are simply connected, which is suitable to evaluate the capability of our method for 2D segmentation. Among $10015$ pairs of images and masks in Ham10000, we divide them into $9000$ pairs for training and $1015$ pairs for testing. All images are downsampled to $128\times 128$ in both the training and testing process. KiTS21 provides segmentation masks for two instances of Kidney annotation, where the connectedness of individuals is strictly preserved. Restricting one continuous mask region is a reliable and practical prior and post-processing strategy. Amongst $300$ cases, we employ $210$ cases for training and the other $70$ cases for testing. All the masks and volume images are normalized into the same spacing and downsampled into a size of $64\times 128\times 128$ after the center is cropped into the same depth. The intensity of images is normalized to $[0,1]$. 

\noindent\textbf{Resources and Parameters}
All models are trained for $300$ epochs with a learning rate of $0.00001$ using RMSprop optimizer. The model for 2-dimensional image segmentation is trained with a minibatch of 64 on a CentOS 7 central cluster computing node with one 64GB, 2.4GHz Intel Xeon E5-2680 CPU, and one GeForce GTX 1080 Ti GPU. 3-dimensional volume image segmentation models are trained with a minibatch of 8 images on a node with one 2.4GHz Intel Xeon E5-2680 CPU, and eight GeForce GTX 1080 Ti GPU. The weighting parameters $\lambda_{Dice}$, $\lambda_{dice}$ and $\lambda_{jac}$ are set to be $1.0$,$1.0$ and $0.1$ respectively.

\subsection{Experiments on 2D Segmentation}
\begin{minipage}{\textwidth}
\centering
    \begin{minipage}[t]{0.45\textwidth}
    \small
    \centering
    \begin{tabular}{@{\hspace*{1mm}}c@{\hspace*{2mm}}c@{\hspace*{2mm}}c@{\hspace*{1mm}}}
    	\toprule
    	Method & Dice & Best \\ \hline
    	UNet&       $93.50 \pm 0.29$    & $93.79$\\ 
    	UNet(cca) & $93.63 \pm 0.29$    & $93.92$\\ 
    	TEDSNet &  $89.91 \pm 0.27$    & $90.18$\\ 
    	TPSN &      $93.77 \pm 0.37$    & $94.14$\\ 
    	mlTPSN&     $94.42 \pm 0.39$    & $94.81$\\
    	\bottomrule
    \end{tabular}
    \captionof{table}{Result on Ham10000}
    \label{tb:LesionBoundary}
    \end{minipage}
    \begin{minipage}[t]{0.45\textwidth}
    \small
    \centering
    \begin{tabular}{@{\hspace*{1mm}}c@{\hspace*{2mm}}c@{\hspace*{2mm}}c@{\hspace*{1mm}}}
    	\toprule
    	Method & Dice & Best\\ \hline
    	UNet&       $92.76 \pm 0.38$ & $93.14$ \\ 
    	UNet(cca) & $92.81 \pm 0.38$ & $93.19$ \\ 
    	TEDSNet &  $87.67 \pm 0.29$ & $88.06$ \\ 
    	TPSN &      $92.73 \pm 0.48$ & $93.21$ \\ 
    	mlTPSN&     $93.17 \pm 0.41$ & $93.58$ \\
    	\bottomrule
    \end{tabular}
    \captionof{table}{Result on KiTS21}
    \label{tb:KiTS21}
    \end{minipage}
\end{minipage}
\medskip

\begin{figure}[t]
    \centering
    \includegraphics[width = \textwidth]{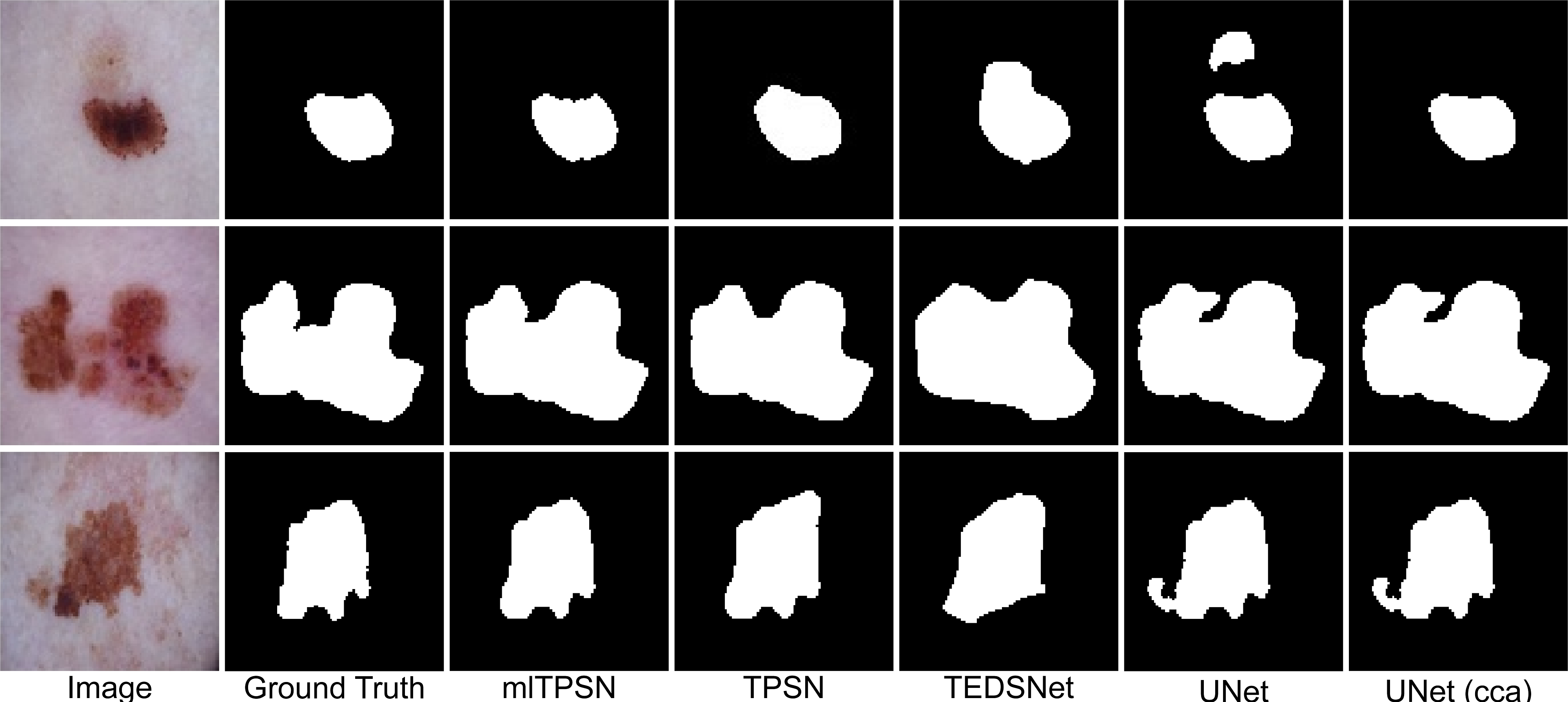}
    \caption{Segmentation results comparison for Ham10000}
    \label{fig:LesionBoundary}
\end{figure}
We conduct experiments on the Ham10000 dataset to validate the capability of TPSN on labeling continuous masks for 2D images. The results are reported in Table.\ref{tb:LesionBoundary}. Compared to UNet segmentation model based on a pixel-wise classification, our methods achieve a better result with $94.14\%$ best Dice. Moreover, mlTPSN achieve a significantly better score with $1.02\%$ higher by Dice ($93.79\%$ to $94.81\%$) than UNet. Using connected component analysis to correct the topological error, the scores are improved to $93.92\%$ by Dice score for the baseline UNet. The results produced by TEDSNet can give a correct topological structure. However, they are not precise and obtain only $90.18\%$ best Dice score.

Figure \ref{fig:LesionBoundary} shows the qualitative visualization. Both TPSN and mlTPSN can avoid topological errors in the first row. As for the results in the second and third rows, the boundaries of the masks predicted by mlTPSN are evidently closer to the ground truth. It demonstrates the effectiveness of the multi-level strategy.

\subsection{Experiments on 3D Segmentation}
We also test the TPSN and ml-TPSN on KiTS21 dataset to segment the 3D kidney. The experimental results are reported in this subsection.
\begin{figure}[htb]
    \centering
    \includegraphics[width = 0.8\textwidth]{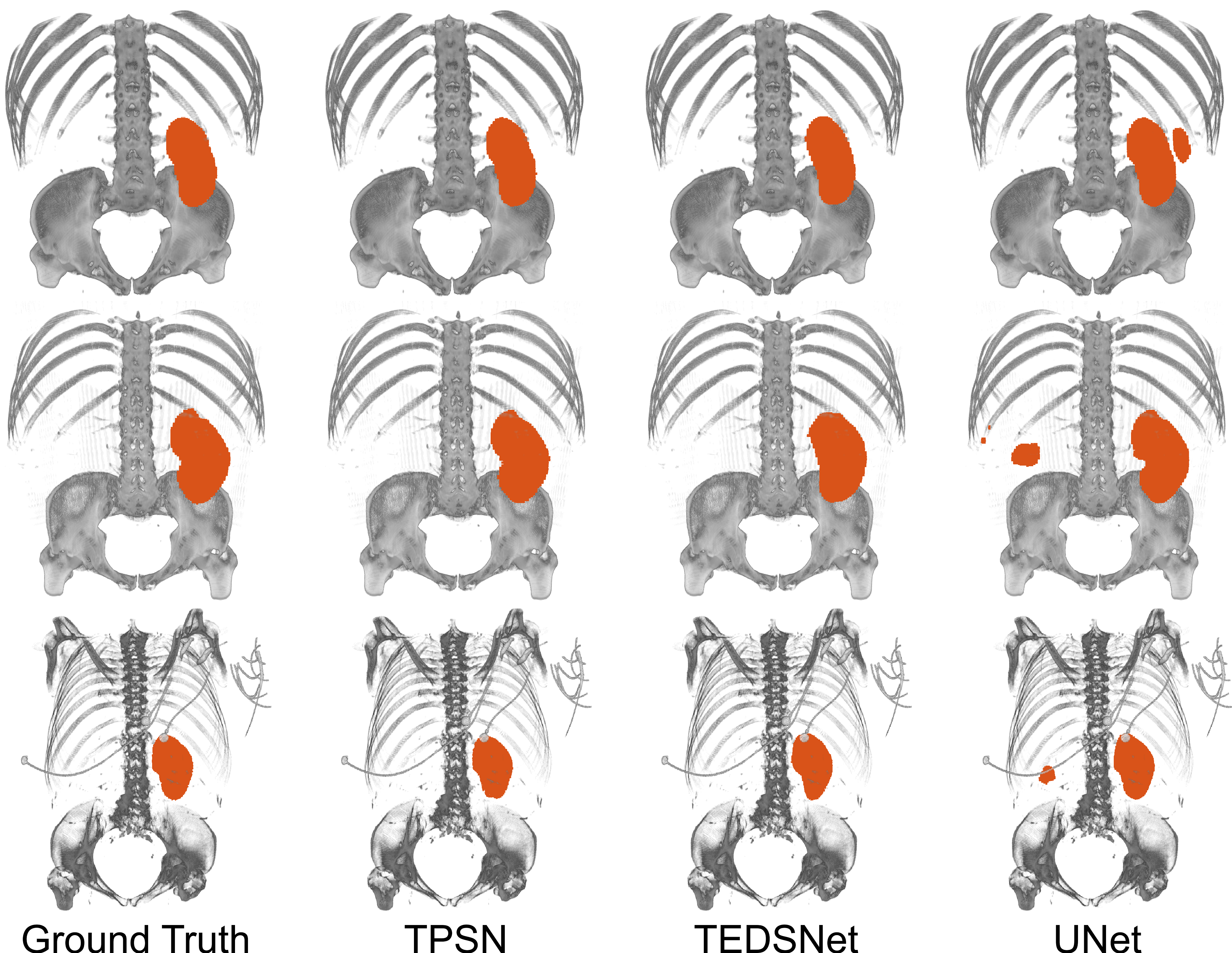}
    \caption{3D Segmentation comparison for KiTS21}
    \label{fig:KiTS21}
\end{figure}

Figure \ref{fig:KiTS21} shows the qualitative visualization results. The masks predicted by our methods are free of outliers and topological errors. On the contrary, the UNet segmentation model mislabels the right kidney while the task is to segment the left one. The quantitative results are reported in Table \ref{tb:KiTS21}. TPSN achieves a higher score by Dice score, which is $93.21\%$. Furthermore, our mlTPSN achieves the best result with $93.62\%$ by Dice. It outperforms all other approaches, including the baseline UNet, UNet with connected component analysis, and TEDSNet.
\begin{figure}[htb]
    \centering
    \includegraphics[width = \textwidth]{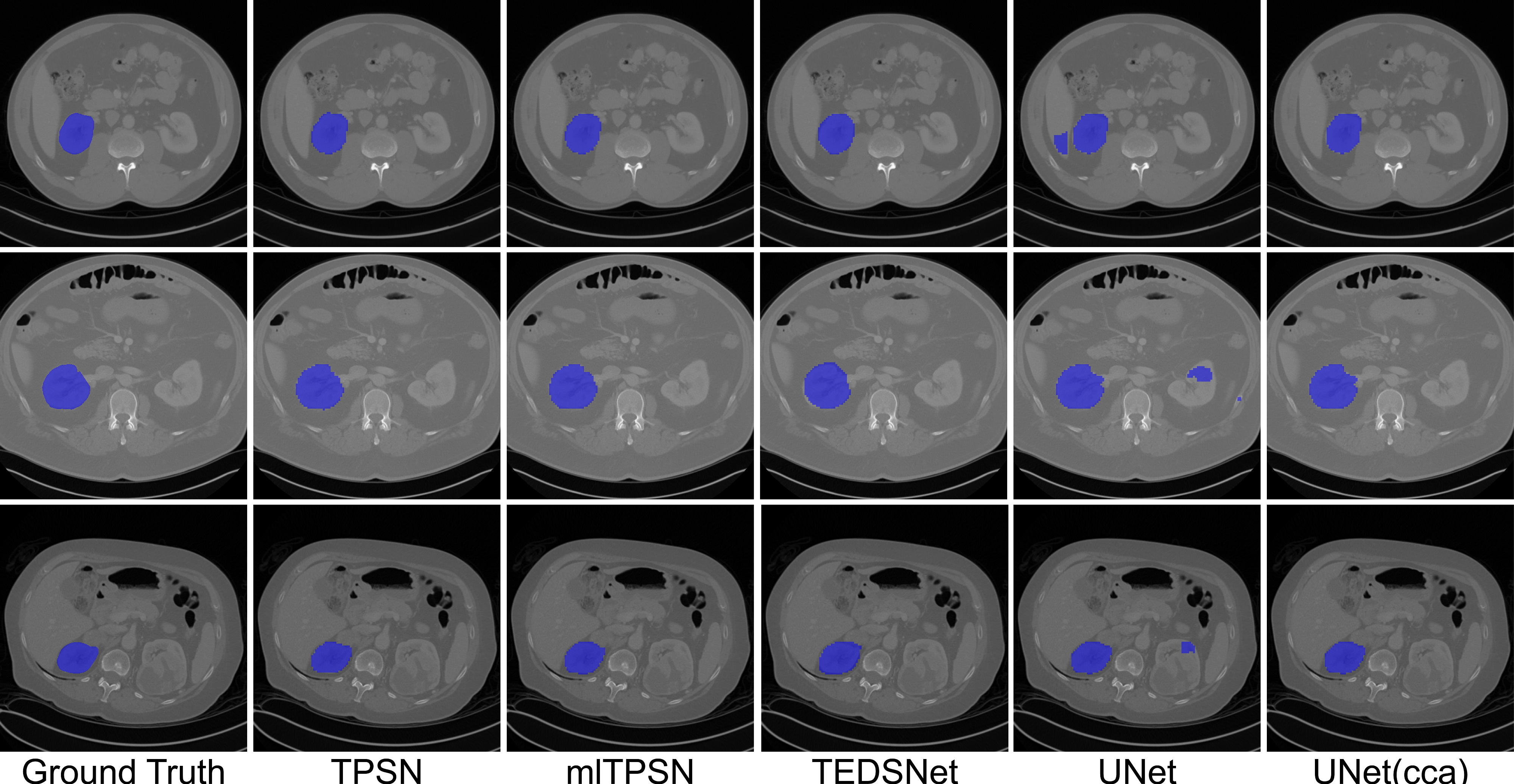}
    \caption{Segmentation results comparison for KiTS21}
    \label{fig:kitsplane}
\end{figure}
\begin{figure}[htb]
    \centering
    \includegraphics[width = \textwidth]{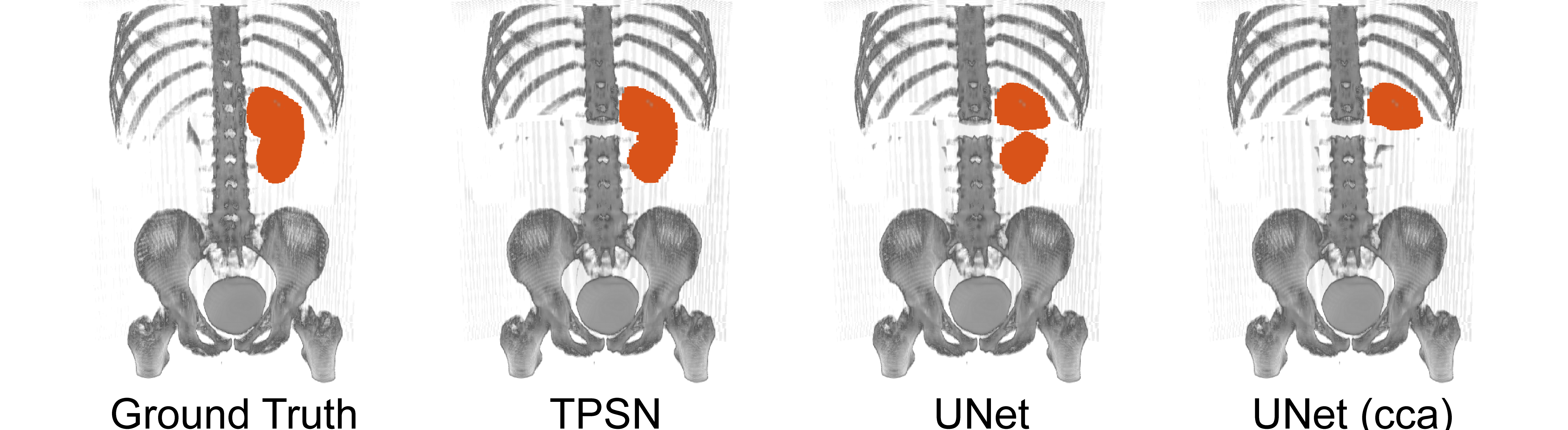}
    \caption{$3D$ Segmentation results comparison for special cases}
    \label{fig:interruption}
\end{figure}

\noindent \textbf{Dataset with Information Loss} We also test our models on images with corruption and information loss. This may be brought by patients' movements or device faults of the scanner. In this experiment, we assume the intensity information has been lost for some slices in the 3D image. It is simulated by setting the intensity value of some slices in the volume image of a subject to be zero. The corrupted image is then fitted into the trained network without any further processing. The result is illustrated in Figure \ref{fig:interruption}. It is observed that the result obtained by TPSN is free of topological errors, even though the mask is shrunk a bit compared to the ground truth. On the contrary, both UNet and UNet(cca) fail to give accurate results. It demonstrates the robustness of our proposed models.
\section{Conclusion}
In this work, we proposed a novel learning-based segmentation framework for both 2D and 3D images that guarantees to preserve the prescribed topology. The main idea is to train a UNet that predicts a diffeomorphic mapping, which registers a template mask to the ground truth. Experimental results show that our methods outperform other methods. Our framework can also be easily incorporated with other architecture to obtain better results.

\bibliographystyle{splncs04}  
\bibliography{reference}

\begin{thebibliography}{10}
\providecommand{\url}[1]{\texttt{#1}}
\providecommand{\urlprefix}{URL }
\providecommand{\doi}[1]{https://doi.org/#1}

\bibitem{chan2018topology}
Chan, H.L., Yan, S., Lui, L.M., Tai, X.C.: Topology-preserving image
  segmentation by beltrami representation of shapes. Journal of Mathematical
  Imaging and Vision  \textbf{60}(3),  401--421 (2018)

\bibitem{chan2001active}
Chan, T.F., Vese, L.A.: Active contours without edges. IEEE Transactions on
  image processing  \textbf{10}(2),  266--277 (2001)

\bibitem{cciccek20163d}
{\c{C}}i{\c{c}}ek, {\"O}., Abdulkadir, A., Lienkamp, S.S., Brox, T.,
  Ronneberger, O.: 3d u-net: Learning dense volumetric segmentation from sparse
  annotation. In: Proc. of Int. Conf. on Medical Image Computing and
  Computer-Assisted Intervention. pp. 424--432. Springer (2016)

\bibitem{heller2021state}
Heller, N., Isensee, F., Maier-Hein, K.H., Hou, X., Xie, C., Li, F., Nan, Y.,
  Mu, G., Lin, Z., Han, M., et~al.: The state of the art in kidney and kidney
  tumor segmentation in contrast-enhanced ct imaging: Results of the kits19
  challenge. Medical Image Analysis  \textbf{67},  101821 (2021)

\bibitem{hu2019topology}
Hu, X., Li, F., Samaras, D., Chen, C.: Topology-preserving deep image
  segmentation. Proc. of Int. Conf. on Neural Information Processing Systems
  (2019)

\bibitem{isensee2021nnu}
Isensee, F., Jaeger, P.F., Kohl, S.A., Petersen, J., Maier-Hein, K.H.: nnu-net:
  A self-configuring method for deep learning-based biomedical image
  segmentation. Nature Methods  \textbf{18}(2),  203--211 (2021)

\bibitem{jha2020doubleu}
Jha, D., Riegler, M.A., Johansen, D., Halvorsen, P., Johansen, H.D.:
  Doubleu-net: A deep convolutional neural network for medical image
  segmentation. In: 2020 IEEE 33rd International symposium on computer-based
  medical systems (CBMS). pp. 558--564. IEEE (2020)

\bibitem{kass1988snakes}
Kass, M., Witkin, A., Terzopoulos, D.: Snakes: Active contour models.
  International journal of computer vision  \textbf{1}(4),  321--331 (1988)

\bibitem{milletari2016v}
Milletari, F., Navab, N., Ahmadi, S.A.: V-net: Fully convolutional neural
  networks for volumetric medical image segmentation. In: Proc. of Int. Conf.
  on 3D Vision (3DV). pp. 565--571. IEEE (2016)

\bibitem{ronneberger2015u}
Ronneberger, O., Fischer, P., Brox, T.: U-net: Convolutional networks for
  biomedical image segmentation. In: Proc. of Int. Conf. on Medical Image
  Computing and Computer-Assisted Intervention. pp. 234--241. Springer (2015)

\bibitem{shit2021cldice}
Shit, S., Paetzold, J.C., Sekuboyina, A., Ezhov, I., Unger, A., Zhylka, A.,
  Pluim, J.P., Bauer, U., Menze, B.H.: cldice-a novel topology-preserving loss
  function for tubular structure segmentation. In: Proc. of IEEE Conf. on
  Computer Vision \& Pattern Recognition. pp. 16560--16569 (2021)

\bibitem{siu2020image}
Siu, C.Y., Chan, H.L., Ming~Lui, R.L.: Image segmentation with partial
  convexity shape prior using discrete conformality structures. SIAM Journal on
  Imaging Sciences  \textbf{13}(4),  2105--2139 (2020)

\bibitem{tschandl2020human}
Tschandl, P., Rinner, C., Apalla, Z., Argenziano, G., Codella, N., Halpern, A.,
  Janda, M., Lallas, A., Longo, C., Malvehy, J., et~al.: Human-computer
  collaboration for skin cancer recognition. Nature Medicine  \textbf{26}(8),
  1229--1234 (2020)

\bibitem{tschandl2018ham10000}
Tschandl, P., Rosendahl, C., Kittler, H.: The ham10000 dataset, a large
  collection of multi-source dermatoscopic images of common pigmented skin
  lesions. Scientific Data  \textbf{5}(1), ~1--9 (2018)

\bibitem{wyburd2021teds}
Wyburd, M.K., Dinsdale, N.K., Namburete, A.I., Jenkinson, M.: Teds-net:
  Enforcing diffeomorphisms in spatial transformers to guarantee topology
  preservation in segmentations. In: Proc. of Int. Conf. on Medical Image
  Computing and Computer-Assisted Intervention. pp. 250--260. Springer (2021)

\bibitem{zhang2021topology}
Zhang, D., Lui, L.M.: Topology-preserving 3d image segmentation based on
  hyperelastic regularization. Journal of Scientific Computing  \textbf{87}(3),
   1--33 (2021)

\bibitem{zhang2021topoconv}
Zhang, D., Tai, X.c., Lui, L.M.: Topology-and convexity-preserving image
  segmentation based on image registration. Applied Mathematical Modelling
  \textbf{100},  218--239 (2021)

\bibitem{zhou2019prior}
Zhou, Y., Li, Z., Bai, S., Wang, C., Chen, X., Han, M., Fishman, E., Yuille,
  A.L.: Prior-aware neural network for partially-supervised multi-organ
  segmentation. In: Proc. of Int. Conf. on Computer Vision. pp. 10672--10681
  (2019)

\end{thebibliography}

\end{document}